\pdfoutput=1
\documentclass[iop,twocolumn,jphysc,8pt,showpacs,floatfix,nofootinbib,
superscriptaddress
]{revtex4-1}
\usepackage{subfigure}
\usepackage{amssymb}
\usepackage{amsfonts}
\usepackage{amsmath}
\usepackage{amsthm}
\usepackage{epsfig}
\usepackage{graphicx}
\usepackage[usenames,dvipsnames]{color}
\usepackage[latin1]{inputenc}
\usepackage{hyperref}
\usepackage{comment}
\begin{document}

\title{Physics-based derivation of a formula for the mutual depolarization of two post-like field emitters}

\author{Fernando F. Dall'Agnol}
\email{fernando.dallagnol@ufsc.br}
\address{Department of Exact Sciences and Education (CEE), Universidade Federal de Santa Catarina,
 Campus Blumenau, Rua Jo\~{a}o Pessoa, 2514, Velha, Blumenau 89036-004, SC, Brazil}

\author{Thiago A. de Assis}
\email{thiagoaa@ufba.br}
\address{Instituto de F\'{\i}sica, Universidade Federal da Bahia,
   Campus Universit\'{a}rio da Federa\c c\~ao,
   Rua Bar\~{a}o de Jeremoabo s/n,
40170-115, Salvador, BA, Brazil}

\author{Richard G. Forbes}
\email{r.forbes@trinity.cantab.net}
\address{Advanced Technology Institute \& Department of Electrical and Electronic Engineering, University of Surrey, Guildford, Surrey GU2 7XH, UK}

\begin{abstract}

Recent analyses of the field enhancement factor (FEF) from multiple emitters have revealed that the depolarization effect is more persistent with respect to the separation between the emitters than originally assumed. It has been shown that, at sufficiently large separations, the fractional reduction of the FEF decays with the inverse cube power of separation, rather than exponentially. The behavior of the fractional reduction of the FEF encompassing both the range of technological interest $0<c/h\lesssim5$ ($c$ being the separation and $h$ is the height of the emitters) and $c\rightarrow\infty$, has not been predicted by the existing formulas in field emission literature, for post-like emitters of any shape. In this letter, we use first principles to derive a simple two-parameter formula for fractional reduction that can be of interest for experimentalists to modeling and interpret the FEF from small clusters of emitters or arrays in small and large separations. For the structures tested, the agreement between numerical and analytical data is $\sim1\%$.

\end{abstract}

\maketitle

In field electron emission (FE) analyses, great attention is paid to macroscopic field enhancement factors (FEFs), since a large FEF usually implies good emission properties. Interacting pairs of identical emitters experience a reduction in their apex FEF ($\gamma_2$) when compared to that ($\gamma_1$) for an isolated emitter. Thus, it is possible to define a ($+ve$) fractional reduction   of the apex FEF via the formula  $\rho=\left(\gamma_1 - \gamma_2\right)/\gamma_1$. This parameter $\rho$ corresponds to the quantity ($-\delta$) used elsewhere \cite{RFJAP2016,arxiv2018}.

Recently, Forbes \cite{RFJAP2016} proved that, in the ``floating sphere at emitter-plane potential (FSEPP)" model, at sufficiently large separations $c$ between the sphere centres, the fractional reduction $\rho$ in apex FEF falls off with a power law decay of $c^{-3}$, rather than the exponential decay usually assumed (e.g. \cite{Bonard,Harris2015AIP,Harris2016}) prior to this work. This means that the interaction effect (which is a form of electrostatic depolarization) persists to larger separations than originally assumed. This behavior was further confirmed in our work involving numerical simulations on several other FE systems \cite{JPCM2018}. Later, Ref. \cite{arxiv2018} derived an analytical proof that, for any pair of protruding structures standing on a conducting plane, the fractional apex-FEF reduction should fall off as $c^{-3}$ for sufficiently large separations. Unfortunately, for most systems, this relationship $\rho \sim c^{-3}$ is difficult to observe at distances within the range of FE applications, which is $c/h\lesssim 5$, where $h$ is the total emitter height.

In this letter, we use first principles arguments, to derive a fitting function for $\rho$ that is adequately valid in the approximate range ($1.5 \lesssim c/h < \infty$). Thus, the formula is useful for technological applications but also has the correct physical form at large separations.

\begin{figure}[h!]
\includegraphics [width=8.5cm,height=4.5cm] {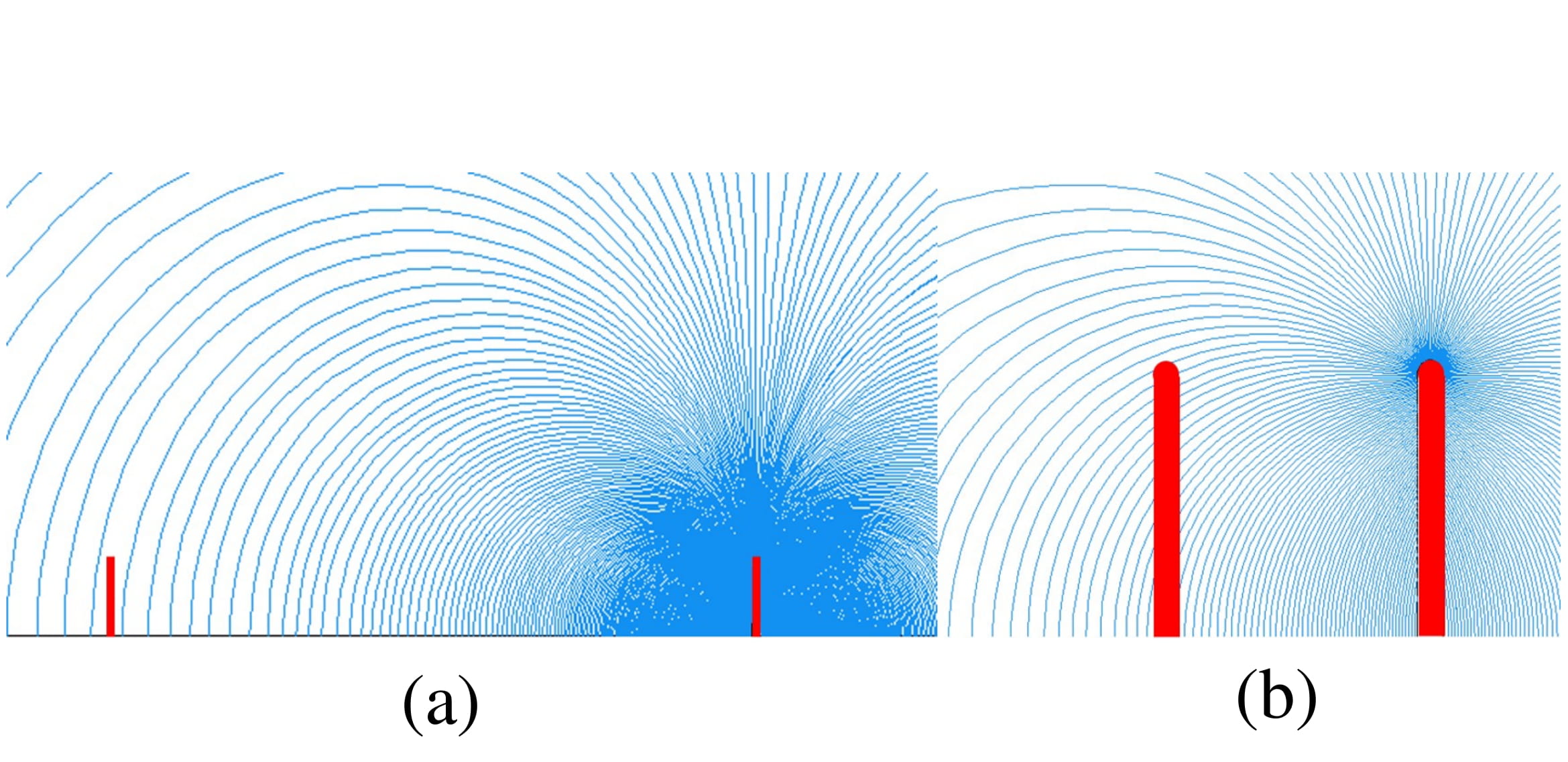}
\caption{Illustration of the depolarizing field. In (a), emitters are far from each other and can be considered dipoles. In (b) their electrostatic field distribution is not uniform at each other position.} \label{Dep}
\end{figure}

Consider two identical conducting posts that stand on a grounded horizontal flat conducting plane and are situated in a vertically uniform, macroscopic field of magnitude $E_M$. Each post develops a charge-distribution near its apex, and there is a corresponding image distribution on the opposite side of the grounded plane. In the lowest order of approximation, these two distributions can each be represented by a finite electrostatic dipole of moment $p$. Thus, from the electrostatic point of view, we are considering two parallel dipoles that tend to mutually depolarize each other. In the lowest order of approximation (which is adequately valid if the posts are sufficiently well separated in comparison with their height), each dipole generates at the midpoint of the other dipole (at the relevant position in the grounded plane) a depolarizing field of magnitude $\Delta E$ given by

\begin{equation}
\Delta E = \frac{p}{4\pi \epsilon_{0} c^{3}},
\label{deltaE}
\end{equation}
where $\epsilon_0$ is the electric constant. If $c$ is sufficiently large, then the depolarizing field strength varies relatively little along the length of the post, and we can define an effective field $E_L$ acting on the post by
\begin{equation}
E_{L} = E_{M} - \Delta E.
\label{EL}
\end{equation}
Figure \ref{Dep} illustrates the depolarizing field that the right-hand side emitter induces on the left. The situation in Fig.\ref{Dep}(a) describes the dipole approximation that we are assuming. At short distances, as in Fig.\ref{Dep}(b), the depolarizing field is not uniform. In this case, we shall attempt to describe the effective  depolarizing field by assuming the polarizability as a variable fitting parameter, as shown below.

The polarization of each emitter is now induced by $E_L$ rather than $E_M$. Hence, as $c$ varies, the extent of mutual depolarization varies. All of $\Delta E$, $E_L$ and $p$ vary with $c$.
By definition of $\alpha$, we can write

\begin{equation}
p = \alpha E_{L},
\label{alpha}
\end{equation}
where $\alpha$ is the effective polarizability of each emitter. Combining Eqs. (\ref{deltaE}) and (\ref{alpha}) in (\ref{EL}), $E_L$ can be written:

\begin{equation}
E_{L} = \frac{E_{M}}{1 + \frac{\alpha}{4\pi \epsilon_{0} c^{3}}}.
\label{EL2}
\end{equation}
By definition, the field at the apex of the emitter when isolated is $E_{1}= \gamma_{1} E_{M}$. The field at the apex of an emitter in a pair is $E_{2}= \gamma_{2} E_{M}$. Equivalently, $E_2$ is also given by $E_2= \gamma_{1}E_{L}$. That is, the field $E_2$ at the apex of a pair of emitters under an applied field $E_M$ is equivalent to an isolated emitter under an applied field $E_{L}$. This implies

\begin{equation}
\gamma_{2} = \frac{\gamma_1}{1 + \frac{\alpha}{4\pi \epsilon_{0} c^{3}}}.
\label{gamma2}
\end{equation}

Finally, the fractional reduction $\rho$ of the apex FEF is

\begin{equation}
\rho = \frac{\gamma_{1}-\gamma_{2}}{\gamma_{1}} = \frac{\alpha}{\alpha + 4\pi \epsilon_{0} c^{3}}.
\label{gamma2rho}
\end{equation}
Equation (\ref{gamma2rho}) predicts the $c^{-3}$ decay at large $c$, as expected, but more importantly, the added term in the denominator leads us to a better fitting function for shorter distances, as we will show below. Note that Eq. (\ref{EL}) is strictly valid only if the depolarizing field is effectively uniform along the post. If the emitters are sufficiently close, then Eq. (\ref{EL}) is not expected to be adequately valid. Nevertheless, the functional form [Eq. (\ref{gamma2rho})] that we have found will lead us to a good fitting function for smaller separations between the emitters, including separations of experimental interest.

\begin{figure}[h!]
\includegraphics [width=8.5cm,height=4.7cm] {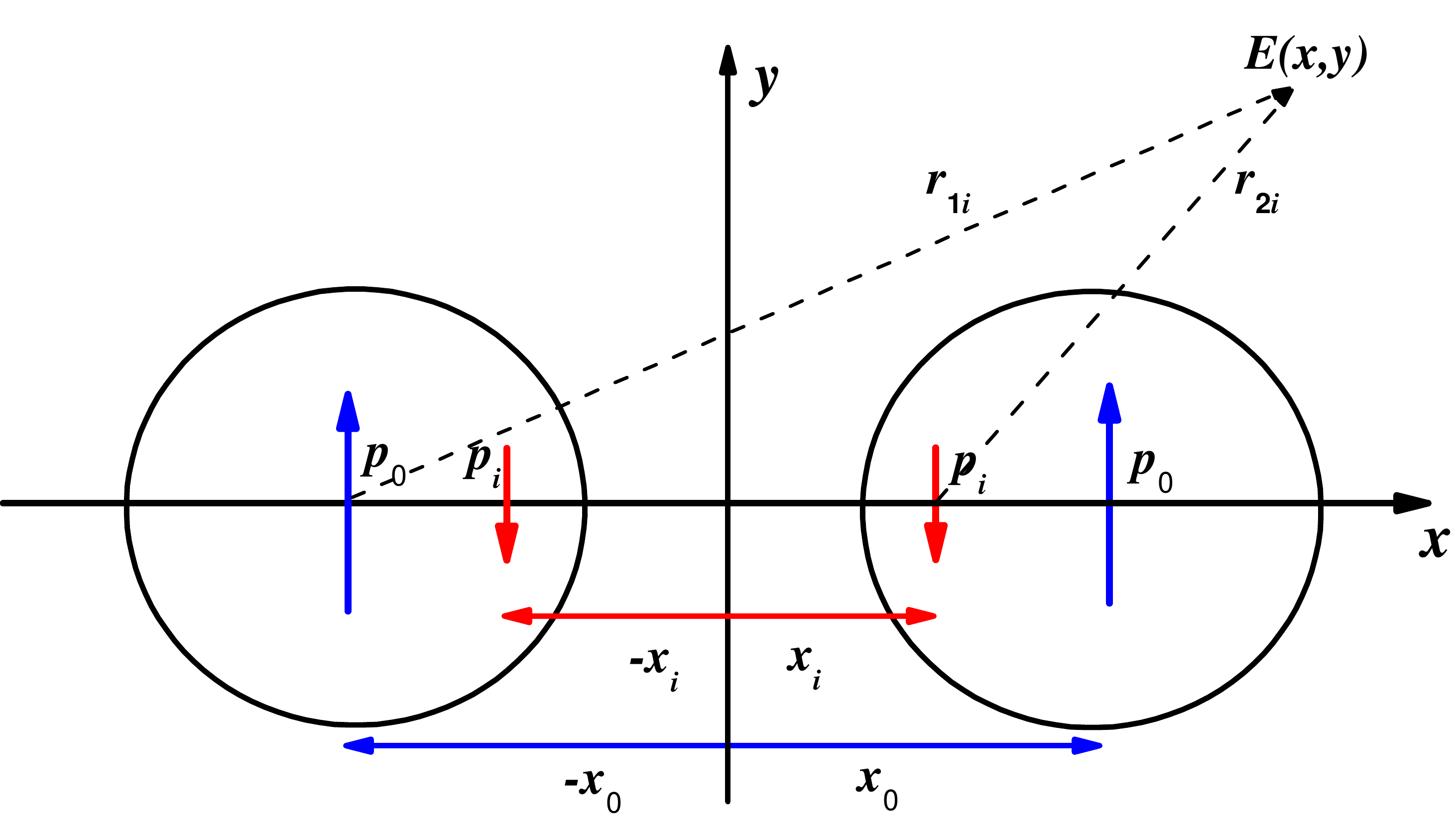}
\caption{Method of images applied to a pair of grounded spheres divided by a grounded plane. The field is computed from the multiple dipole images that each sphere generates in the other. The solution of a pair of hemispheres on a grounded plane is equivalent to the solution in the upper half of this system.} \label{DepFig}
\end{figure}

Next, we will check how well Eq. (\ref{gamma2rho}) can fit $\rho$ in a system with an analytical solution. Consider a physical system consisting of two conducting hemispheres on a conducting plane, all embedded in a vertical uniform electric field. By symmetry, the electric field distribution in this system is the same as for a pair of grounded spheres (Fig.\ref{DepFig}). We employ the method of images to determine the electric field distribution \cite{Bosch,Wallen,RBEF2012}. Hereafter, variables in bold represent a vector, and the corresponding non-bold symbols represent the norm of the vector. Let $p_{0}$ be the dipole moment associated with an isolated conducting sphere in a uniform applied field $\mathbf{E_M}$. As two spheres come closer, the dipole of strength $\mathbf{p_0}$ from each sphere induces an image dipole of strength $\mathbf{p_1}$ in the other, which in turn generates a further image dipole of strength $\mathbf{p_{2}}$, and so on. The $x$-coordinate and the magnitude of the $i$-th dipole are obtained from the following recurrence relations \cite{RBEF2012}:

\begin{equation}
x_{i} = x_{0} - \frac{a^{2}}{x_0 + x_{i-1}},
\label{rec0}
\end{equation}

\begin{equation}
p_{i} = -a \frac{x_{0}-x_{i}}{\left(x_{0} +x_{i}\right)^{2} }p_{i-1},
\label{rec1}
\end{equation}
where $x_0$ is the coordinate of the center of the sphere, $a$ is the radius, and $p_0 = 4\pi\epsilon_{0}a^{3}E_{M}$. The total electric field can be readily calculated from all dipoles plus the applied field, resulting in:

\begin{equation}
\mathbf{E}(x,y) = \mathbf{E_{M}} + \frac{1}{4\pi \epsilon_{0}} \sum\limits_{i=0}^{\infty} \sum\limits_{n=1}^{2} \left[\frac{\frac{3\left(\mathbf{p_{i}}\cdot\mathbf{r_{ni}}\right)\mathbf{r_{ni}}}{r_{ni}^2} - \mathbf{p_{i}}}{r_{ni}^{3}}\right],
\label{rec2}
\end{equation}
where,

\begin{equation}
\mathbf{r_{1i}} = \left(x+x_{i}, y\right),
\label{rec4}
\end{equation}

\begin{equation}
\mathbf{r_{2i}} = \left(x-x_{i}, y\right),
\label{rec5}
\end{equation}

and

\begin{equation}
\mathbf{E_{M}} = \left(0, -E_{M}\right).
\label{rec6}
\end{equation}

The FEF at the apex is defined as

\begin{gather}
\gamma_{apex} = \frac{\mathbf{E}(x_{0},a)}{E_{M}} = \nonumber \\
 = \left|(0,-1) + \frac{1}{4\pi \epsilon_{0}E_{M}} \sum\limits_{i=0}^{\infty} \sum\limits_{n=1}^{2} \left[\frac{\frac{3\left(\mathbf{p_{i}}\cdot\mathbf{r_{ni}}\right)\mathbf{r_{ni}}}{r_{ni}^2} - \mathbf{p_{i}}}{r_{ni}^{3}}\right]\right|.
\label{rec7}
\end{gather}
%

It is convenient to define a ratio  $\xi_{i}= p_{i}/p_{0}$, in order to somewhat simplify Eq. (\ref{rec7}), to make the apex FEF depend only on geometrical parameters. Thus:

\begin{gather}
\gamma_{apex} = \nonumber \\
= \left|(0,-1) + a^{3} \sum\limits_{i=0}^{\infty} \left[3a\xi_{i} \left(\frac{\mathbf{r_{1i}}}{r_{1i}^{5}} +\frac{\mathbf{r_{2i}}}{r_{2i}^{5}}\right) - \left(\frac{1}{r_{1i}^{3}} + \frac{1}{r_{2i}^{3}}\right)\mathbf{\xi_i}\right]\right|,
\label{rec8}
\end{gather}
where  $\mathbf{\xi_{i}} =(0, \xi_{i})$ and the recurrence relation for $\xi_i$ is the same as for $p_i$ in Eq. (\ref{rec1}). From Eq. (\ref{rec7}) to (\ref{rec8}) we used $\mathbf{p_i}=(0,p_{0}\xi_i)$; $\mathbf{r}_{1i}|_{apex}=(x_0+x_i,a)$; $\mathbf{r}_{2i}|_{apex}=(x_0-x_i,a)$. Then, $\mathbf{p}i\cdot\mathbf{r}_{1i}= \mathbf{p}i\cdot\mathbf{r}_{2i}= p_{0}a\xi_{i}= 4\pi\epsilon_{0}a^{4}E_{M}\xi_{i}$.

Having defined $\gamma_{apex}$, the FEF fractional reduction is

\begin{equation}
\rho_{an}=\frac{\gamma_1 -\gamma_{apex}}{\gamma_1}= 1 - \frac{\gamma_{apex}}{3}.
\label{rhoannn}
\end{equation}

\begin{figure}[h!]
\includegraphics [width=7.5cm,height=5.7cm] {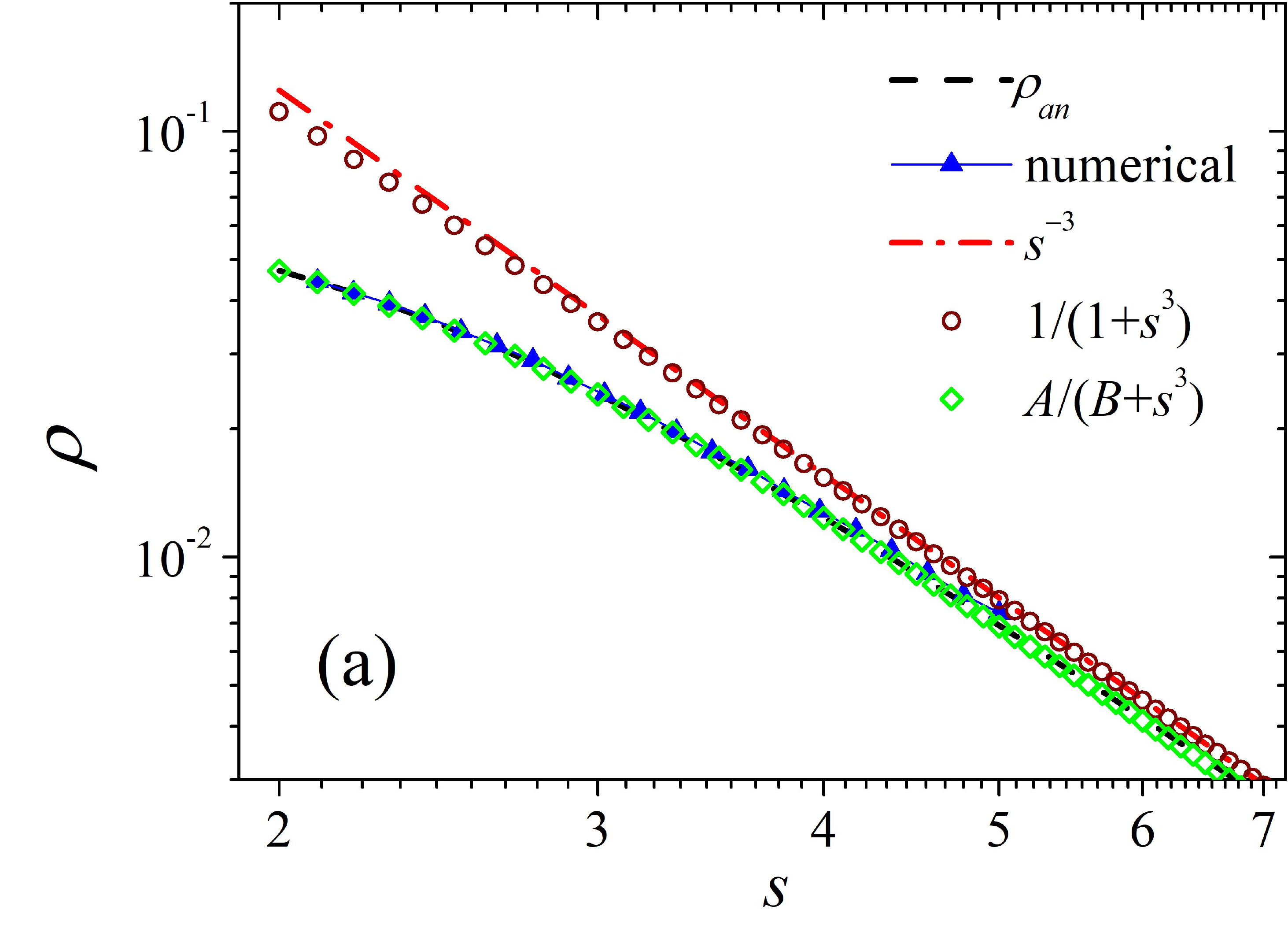}
\includegraphics [width=7.5cm,height=5.7cm] {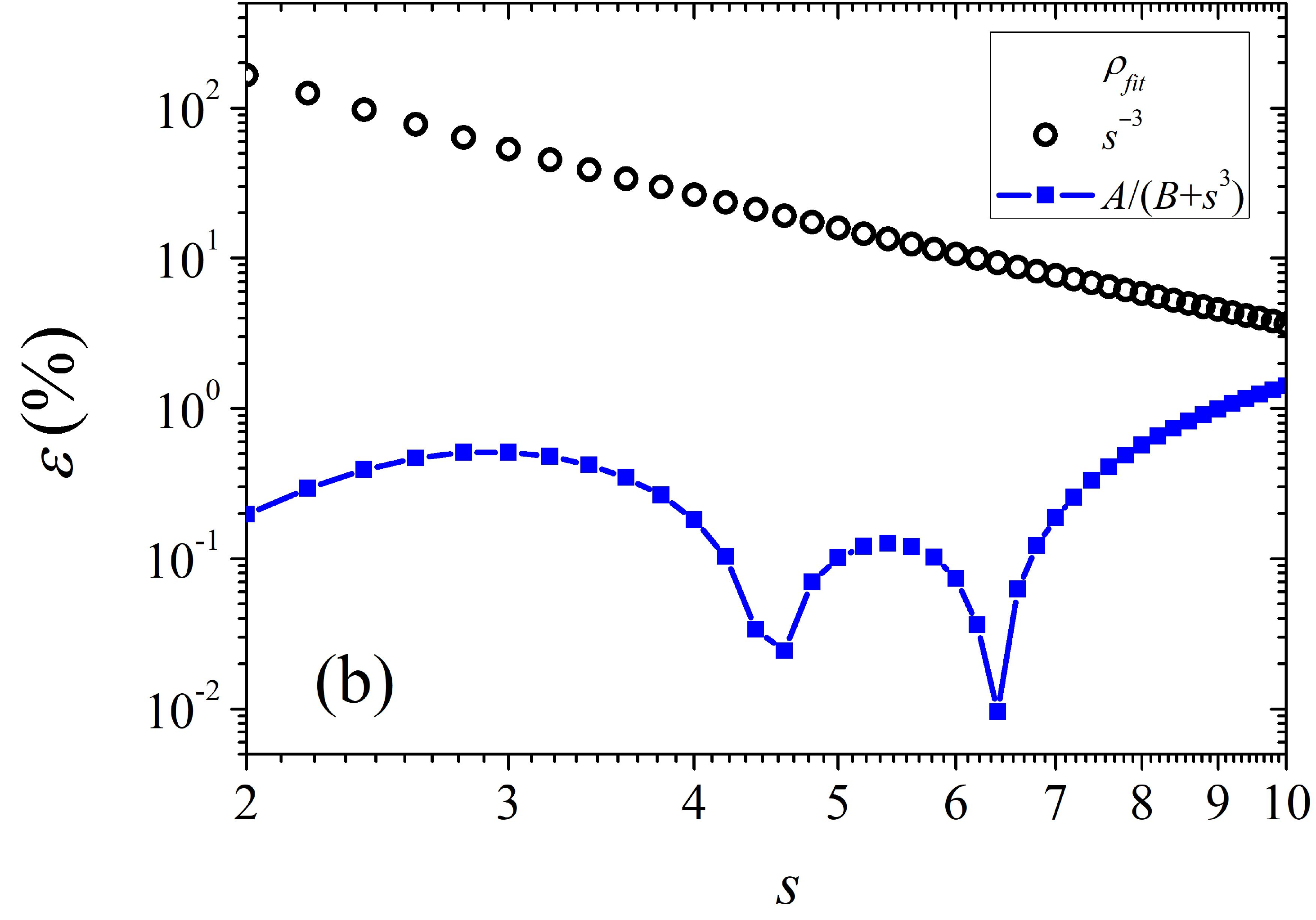}
\caption{(a) Comparison, for the two-sphere case, between the exact analytical result ($\rho_{an}$), numerical results and several candidate fitting functions. The fitting function we propose shows excellent agreement when $A=(0.9433\pm0.0006)$ and $B=(12.00\pm0.02)$, such that the superposition with $\rho_{an}$ is almost indistinguishable on this scale. (b) Relative difference between our fitting function and $\rho_{an}$ which is less than 1\% over a large range including the range of technological interest.} \label{DepFigRes}
\end{figure}

The label ``$an$" in $\rho_{an}$ stands for ``analytical". The black full line and the curve in blue triangles in Fig.\ref{DepFigRes}(a) shows $\rho_{an}$ compared to a numerical simulation based on finite elements to confirm of our result.

The value of  $\gamma_{apex}(x_{0}\rightarrow \infty)$ for sufficiently large separations, can be determined by considering only the first term in the summation of Eq. (\ref{rec8}), $x=x_0$ and $y=a$, which results in

\begin{gather}
\gamma_{apex}(x_{0}\rightarrow \infty) = \nonumber \\
= \left|(0,-3) - \frac{3a^{4}(2x_{0},a)}{\left(4x_{0}^{2} + a^2\right)^{5/2}} + \frac{a^{3}(0,1)}{\left(4x_{0}^{2} + a^2\right)^{3/2}} \right|.
\label{rec8}
\end{gather}

The middle term in Eq. (\ref{rec8}) tends to zero faster than the last term and can be ignored. The resulting vector has only the $y$-component and the norm simplifies to

\begin{equation}
\gamma_{large} = 3 - \frac{a^{3}}{8x_{0}^{3}}.
\label{gammalarge}
\end{equation}

\begin{figure}[h!]
\includegraphics [width=7.5cm,height=5.7cm] {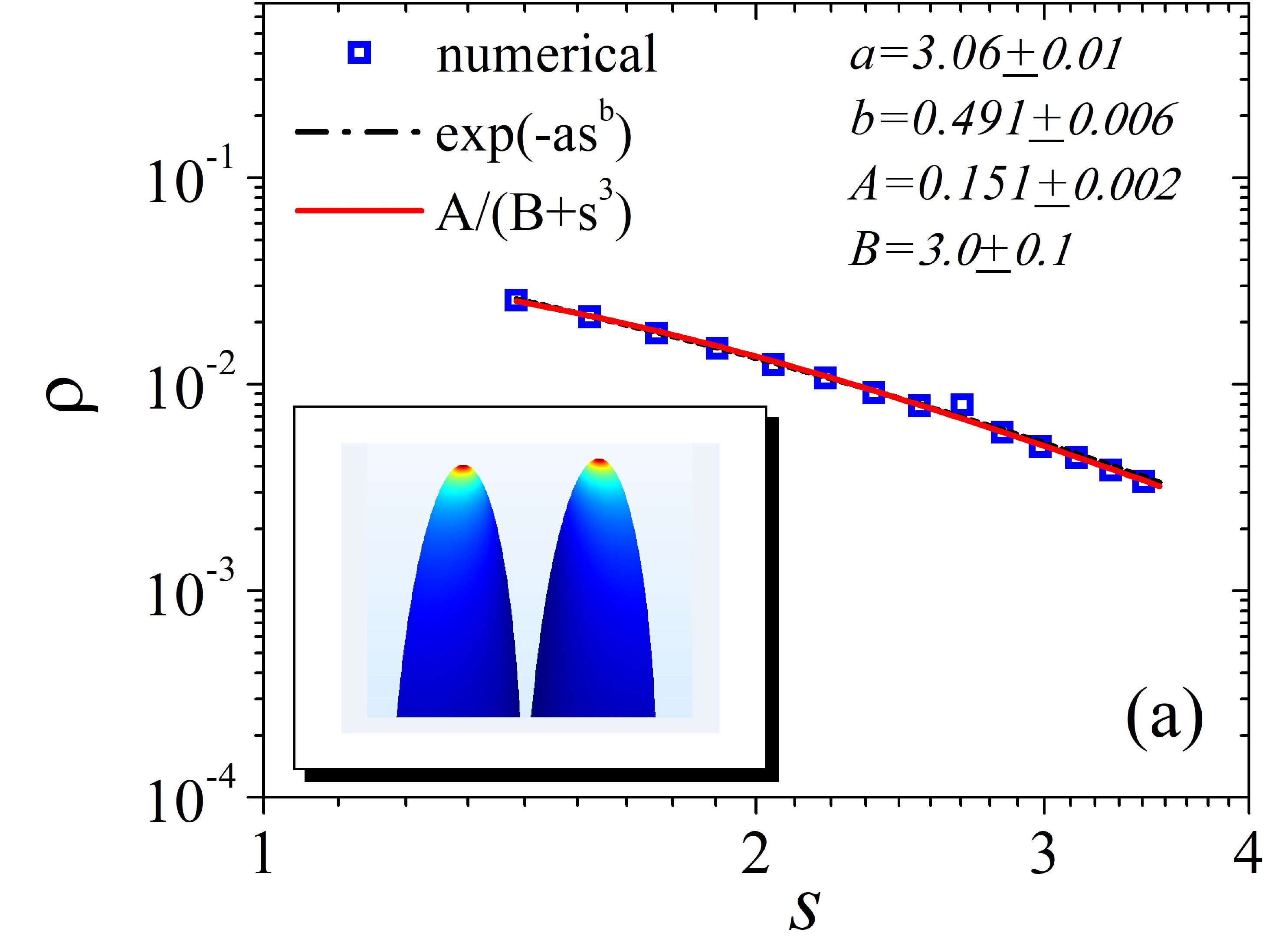}
\includegraphics [width=7.5cm,height=5.7cm] {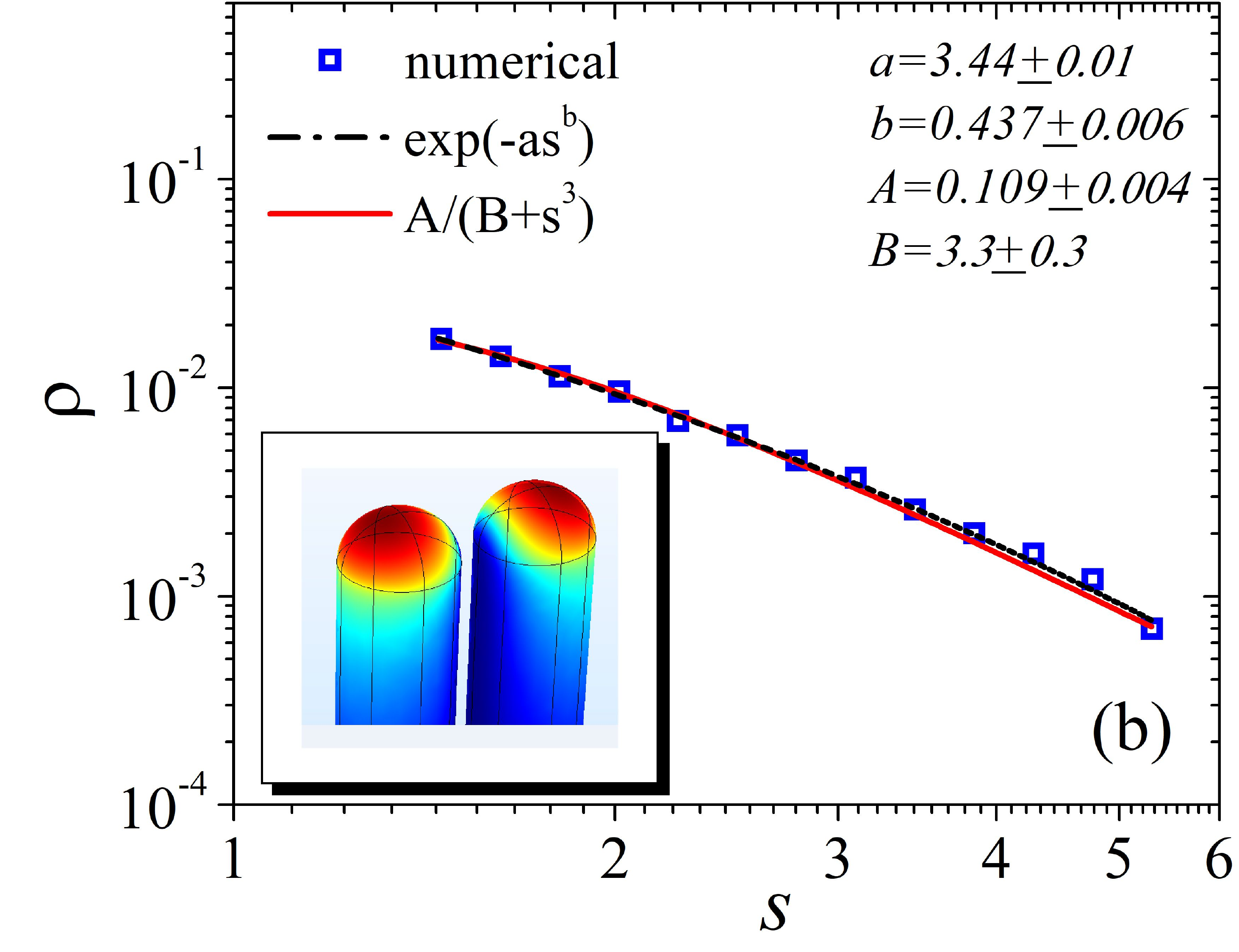}
\includegraphics [width=7.5cm,height=5.7cm] {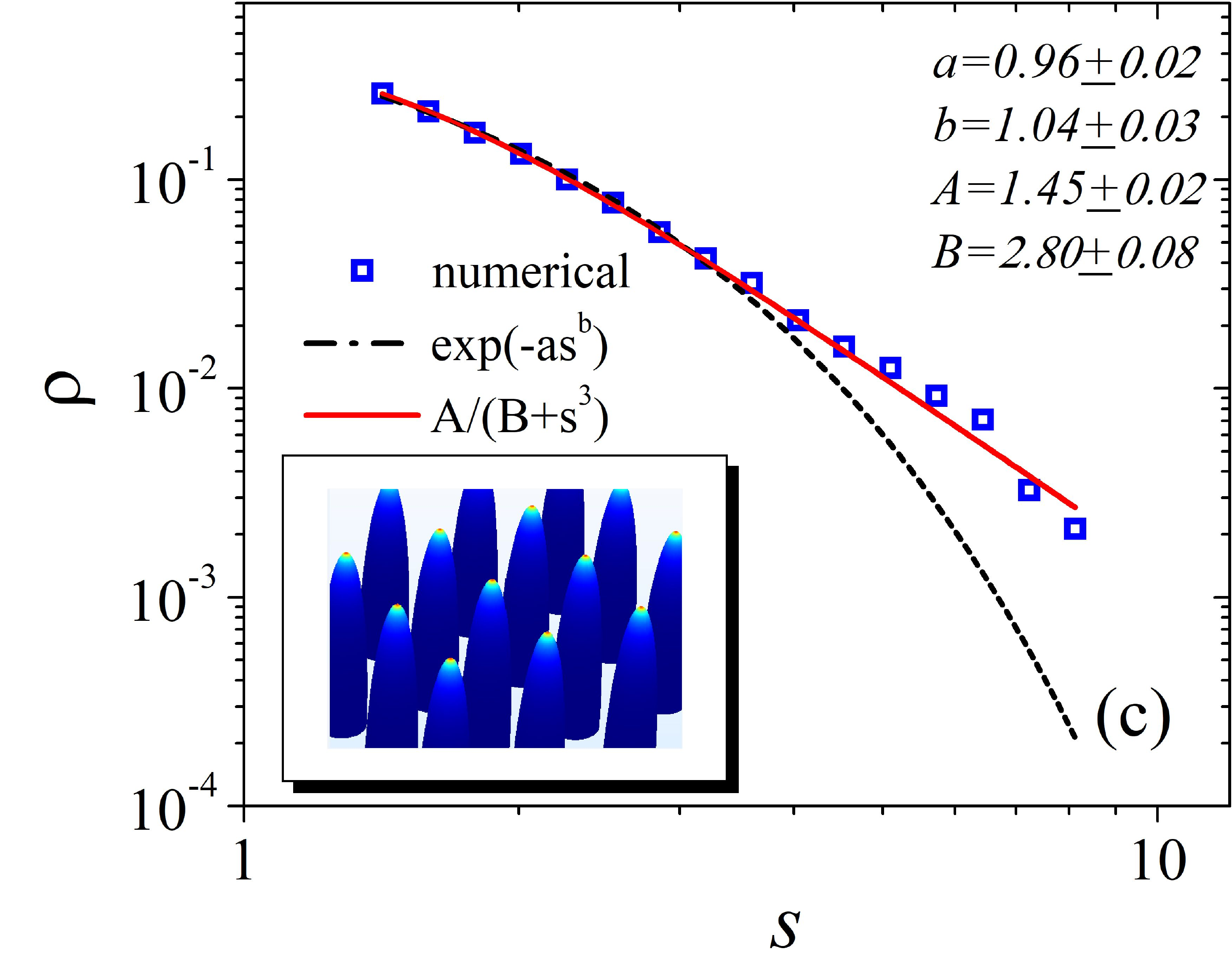}
\caption{Numerical results for three distinct systems fitted with a formula from Harris \textit{et al.} \cite{Harris2015AIP,Harris2016} and with our Eq. (\ref{gammafit}). The latter shows very good agreement in all range of distances considered. System (a) consists of a pair of identical ellipsoids, (b) consists of a pair of identical HCP emitters and (c) shows an infinite array of ellipsoids. In systems (a) and (c) each emitter's aspect ratio is $h/r=8$. In system (b) each emitter's aspect ratio is $h/r=51$. The fittings were performed by a least squares method. The color map in the insets represents the local FEF [red (blue) color indicates higher
(lower) local FEF].} \label{Systems4}
\end{figure}

From Eq. (\ref{rec8}) it is obvious that $\gamma_{large}\rightarrow \gamma_{1}=3$ (isolated emitter) as $x_{0}\rightarrow\infty$, as expected. Then $\rho_{large}=\left(3 - \gamma_{large}\right)/3$, for sufficiently large distances is

\begin{equation}
\rho_{large} = \frac{a^{3}}{c^{3}} = \frac{1}{s^{3}},
\label{gammalarge2}
\end{equation}
where $c=2x_{0}$ (distance between emitters) and $s$ is a spacing parameter defined as $c$ normalized by the height of the emitter ($s=c/a$). This tidy expression is exact for two hemispheres and provide an analytical proof of a power-law decay with power $-3$, reinforcing the universality reported in Ref.\cite{JPCM2018}.

The exact analytical result $\rho_{an}$ and the simple approximation $\rho_{large}$ converge for sufficiently large $s$. A criterion for what is ``sufficiently large" can be defined as $\varepsilon =\left(|\rho_{an} -\rho_{large}|\right)/ \rho_{an} \times 100\%$. The value of $\varepsilon$ indicates how far is enough for Eq. (\ref{gammalarge2}) to be able to predict the exact analytical result within $\varepsilon\%$ of accuracy. For example, for $s>5$, the maximum error is $\varepsilon= 16\%$. As another example, a maximum error of 1\% can be achieved only for $s >18$ (very large $s$). These examples show that Eq. (\ref{gammalarge2}) provides a poor general fitting function for $\rho$ since the range of theoretical and experimental interest is about $0< s < 5$.

With Eq. (\ref{gamma2rho}), knowing that $\rho(s\rightarrow\infty)=1/s^3$ imposes the condition $\alpha/4 \pi \epsilon_0a^3=1$. Then, plotting  $\rho=1/(1+s^3)$, shows a little improvement towards $\rho_{an}$, although still unsatisfying [see Fig. \ref{DepFigRes}(a)]. We know that Eq. (\ref{gamma2rho}) has a limitation at small $s$, because $E_L$ becomes non-uniform along the post. Hence, we attempted to compensate whatever effect the non-uniformity causes by turning the constants in the expression for $\rho$ into fitting parameters as in

\begin{equation}
\rho_{fit} = \frac{A}{B+s^{3}}.
\label{gammafit}
\end{equation}

Now, the plot of Eq. (\ref{gammafit}) for $A=(0.9433\pm0.0006)$ and $B=(12.00\pm0.02)$, values found using least squares method, is satisfactorily close to $\rho_{an}$ in all relevant ranges of $s$. Figure \ref{DepFigRes}(a) shows the improvement in $\rho$  from  $\rho_{large}$ to  $\rho_{an}$ as $\rho_{fit}(A,B)$ is switched from  $\rho_{fit}(1,1)$ to $\rho_{fit}(0.9433\pm0.0006, 12.00\pm0.02)$. In the latter case, $\rho_{an}$ and $\rho_{fit}$ are almost superimposed, on the scale used. Figure \ref{DepFigRes}(b) shows the relative error $\varepsilon$ in each case.

It should be noted that approximation (\ref{gammafit}) may fail for very small $s$-values, certainly when $s \lesssim0.05$. In this case, the functional dependence of the fractional reduction $\rho$ will be significantly affected by additional electrostatic effects that, depending on the geometry of the system, can be related to the ``close proximity electrostatic effect" \cite{Jensen2015,FT2017JPCM}.

The good fitting properties of Eq. (\ref{gammafit}), except at very small $s$-values, have been verified for a large number of different physical systems that have only numerical solutions. As an example, we take the case of (i) a pair of hemi-ellipsoidal emitters, (ii) an infinite array of hemi-ellipsoidal emitters (both (i) and (ii) systems with each emitter's aspect ratio $h/r=8$, where $r$ is the base radius) and (iii) a pair of HCP model (hemisphere on cylindrical post) emitters, each with aspect ratio $h/r = 51$. We compare here the results from the numerical solution of Laplace equation by using the finite-element method \cite{JPCM2018}, with both Eq. (\ref{gammafit}) and an exponential fitting used by Harris \textit{et al.} \cite{Harris2015AIP,Harris2016}. Figure \ref{Systems4} illustrates the results for all systems studied. It is useful to notice that, for fixed $s$, $\rho$ is larger for the infinite array system, as a consequence of stronger depolarization. Also, both fitting formulae show good agreement in a range $1.5\lesssim s \lesssim 3.5$, but Eq. (\ref{gammafit}) has the added benefits that it is informed by electrostatic principles and can predict the physically correct power law decay $c^{-3}$ for large $s$.

A final remark regarding the parameter $A$ of the Eq. (\ref{gammafit}): if we consider the two floating spheres in the FSEPP model described in Ref.\cite{RFJAP2016}, in the limit of large separations one obtain $\rho_{large} \approx 2 r h^2/c^3$. Thus, it is possible to write in this system

\begin{equation}
\rho_{large} \approx 2\times \left(\frac{h}{r}\right)^{-1} \times s^{-3}.
\label{fit222}
\end{equation}

For sufficient large separations (where $B \ll s^{3}$), Eq. (\ref{gammafit}) gives $\rho_{large}\approx A \times s^{-3}$. If we compare the latter approximation and Eq. (\ref{fit222}), it is possible to make the correspondence,

\begin{equation}
A \approx 2\times \left(\frac{h}{r}\right)^{-1}.
\label{corresp}
\end{equation}
Thus, the parameter $A$ is, in general, expected to be dependent on the ratio $h/r$. This is an important result, since it was not reported in the parameters used in the exponential formulas by Harris \textit{et al.} \cite{Harris2015AIP,Harris2016} nor before by Bonnard \textit{et al.} \cite{Bonard} or Jo \textit{et al.} \cite{Jo}. In contrast, Refs.\cite{Bonard,Jo} have assumed a constant coefficient, usually taken as $-2.3172$, in their exponential dependence between $\rho$ and $c/h$. This aspect certainly deserves further investigation.

In summary, we have employed a simple approach, that is informed by electrostatic arguments and a dipole approximation, to determine the effects of mutual charge blunting (i.e., mutual electrostatic depolarization) on a pair of conducting posts. This yields a formula for the fractional reduction in apex field enhancement factor, as a function of post separation. The formula may be useful for experimentalists in field emission related technologies, and also provides insights into the physics of electrostatic interactions in small clusters of emitters, at moderate to large separations ($c/h \gtrsim 1.5$). Although the modeling of depolarization effects at small distances needs to be more sophisticated than the dipole-dipole approach, the functional dependence of $\rho$ found in Eq. (\ref{gamma2rho}) led us to a fitting formula that better describes the physical behavior expected for $\rho$. We compared our fitting formula [Eq. (\ref{gammafit})] with the formula from Harris \textit{et al.} \cite{Harris2015AIP,Harris2016} for several systems. Both formulae perform well at separations of technological interest. However, Eq. (\ref{gammafit}) has the added benefit of being informed by electrostatic principles and can predict the physically correct power law decay $c^{-3}$ for large $s$. The fitting parameters $A$ and $B$ are connected with the non-uniformity of the local electrostatic field $E_L$ along the post and the parameter $A$ also depends on the aspect ratio. By choosing appropriate values of the fitting parameters, we were able to use Eq. (\ref{gammafit}) to predict the behavior of the fractional FEF reduction $\rho$ for a large number of systems containing protrusions of various shapes, in the region of moderate separations (such that $c/h \gtrsim 1.5$), which is the range of technological interest.

TAdA and RGF thank Royal Society financial support under Newton Mobility Grant, Ref: NI160031.  TAdA and FFD also thank CNPq (Brazilian agency).

\providecommand{\newblock}{}

\end{document}